\documentclass[aps,pre,superscriptaddress,nofootinbib,10pt]{revtex4-2}

\usepackage[utf8]{inputenc} 
\usepackage[T1]{fontenc}    
\usepackage{hyperref}       
\usepackage{url}            
\usepackage{booktabs}       
\usepackage{amssymb,amsmath,amsfonts, dsfont}       
\usepackage{nicefrac}       
\usepackage{graphicx}       
\graphicspath{{./grafici/}} 
\usepackage{float}
\makeatletter
\let\newfloat\newfloat@ltx
\makeatother
\usepackage{algorithm} 
\usepackage{algpseudocode} 
\usepackage{physics}
\usepackage[dvipsnames]{xcolor}
\usepackage{enumitem}

\newcommand{\bea}{\begin{eqnarray}}
\newcommand{\eea}{\end{eqnarray}}

\newcommand{\<}{\langle}
\renewcommand{\>}{\rangle}

\newcommand{\HH}{\mathcal{H}}
\newcommand{\PGB}{P_\text{GB}}

\newcommand{\s}{\sigma}
\newcommand{\bs}{\boldsymbol{\sigma}}

\newcommand{\new}[1]{\textcolor{black}{#1}}

\begin{document}

\title{Performance of machine-learning-assisted Monte Carlo\\ in sampling from simple statistical physics models}

\author{Luca Maria Del Bono}
\affiliation{Dipartimento di Fisica, Sapienza Università di Roma, Piazzale Aldo Moro 5, Rome 00185, Italy}
\affiliation{CNR-Nanotec, Rome unit, Piazzale Aldo Moro 5, Rome 00185, Italy}

\author{Federico Ricci-Tersenghi}
\affiliation{Dipartimento di Fisica, Sapienza Università di Roma, Piazzale Aldo Moro 5, Rome 00185, Italy}
\affiliation{CNR-Nanotec, Rome unit, Piazzale Aldo Moro 5, Rome 00185, Italy}
\affiliation{ INFN, sezione di Roma1, Piazzale Aldo Moro 5, Rome 00185, Italy}

\author{Francesco Zamponi}
\affiliation{Dipartimento di Fisica, Sapienza Università di Roma, Piazzale Aldo Moro 5, Rome 00185, Italy}

\begin{abstract}
Recent years have seen a rise in the application of machine learning techniques to aid the simulation of hard-to-sample systems that cannot be studied using traditional methods. Despite the introduction of many different architectures and procedures, a wide theoretical understanding is still lacking, with the risk of suboptimal implementations. As a first step to close this gap, we provide here a complete analytic study of the widely-used Global Annealing (also called Sequential Tempering) procedure applied to a shallow MADE architecture for the Curie-Weiss model. The contribution of this work is twofold: firstly, we give a description of the optimal weights and of the training under Gradient Descent optimization. Secondly, we compare what happens in Global Annealing with and without the addition of local Metropolis Monte Carlo steps. We are thus able to give theoretical \new{insight} into the best procedure to apply in this case. This work establishes a clear theoretical basis for the integration of machine learning techniques into Monte Carlo sampling and optimization.
\end{abstract}

\maketitle

\section{Introduction}

\subsection{\new{Setting and state-of-the-art}}

Obtaining configurations of hard-to-sample systems, such as spin glasses, amorphous solids and proteins, is a challenging task with both theoretical and practical applications \cite{mezard1987spin, 10.23943/princeton/9780691147338.003.0007, charbonneau2023spin}. 
The goal is to sample independent and identically distributed configurations from
the Gibbs-Boltzmann (GB) distribution
\begin{equation}\label{eq:GB_distribution}
    \PGB(\bs) = \frac{e^{-\beta \HH(\bs)}}{\mathcal{Z}(\beta)},
\end{equation}
where $\bs = \{\s_1,\dots,\s_N\}$ is a set of \new{$N$} random variables describing the system, $\mathcal{Z}(\beta)$ is the partition function, $\HH(\bs)$ is
the Hamiltonian 
and $\beta = 1/T$ is the inverse temperature.

Among the most widely used techniques are Parallel Tempering (PT)~\cite{hukushima1996exchange, houdayer2001cluster, zhu2015efficient} and Population Annealing (PA)~\cite{hukushima2003population, machta2010population, wang2015comparing, barash2017gpu, weigel2021understanding}, which have established themselves as state-of-the-art approaches over the past few decades, remaining largely unchanged from their original formulations.
In recent years, a new line of research aims to use machine-learning-assisted techniques to aid sampling. Stemming from the increasing capabilities of Generative Artificial Neural Networks, this line of research aims at using different kinds of architectures and frameworks, such as autoregressive models~\cite{bialas2022hierarchical, ciarella2023machine, biazzo2024sparse, del2024nearest}, normalizing flows~\cite{schonle2023optimizing, nicoli2023detecting}, diffusion models~\cite{biroli2023generative, bae2024very, hunt2024accelerating} and Boltzmann machines~\cite{decelle2024restricted}, to aid the sampling of statistical mechanics models. Methods inspired by theoretical physics techniques such as the Renormalization Group~\cite{marchand2022wavelet, masuki2025generative} have been introduced and non-generative-based techniques can be used as well~\cite{galliano2024policy}.
Moreover, similar techniques have been applied to the related task of finding the lowest energy configuration of statistical mechanics systems, a task with deep practical implications~\cite{pugacheva2024enhancing, shen2025free}.
At the same time, the Neural Network (NN) can also be used to obtain variational approximations to the GB distribution, thus improving over mean-field techniques, both in classical~\cite{wu2019solving,biazzo2024sparse} and quantum~\cite{carleo2017solving} systems.

One of the most widely used NN-based techniques is the Global Annealing (GA) procedure~\cite{mcnaughton2020boosting, ciarella2023machine}\new{, also called \textit{Sequential Tempering} in the literature~\cite{mcnaughton2020boosting}}. In GA, a set of $M$ configurations is progressively cooled to lower temperatures. At each step, starting from $M$ equilibrated configurations, a new series of configurations is generated using a neural network at a slightly lower temperature. The new configurations are equilibrated at the lower temperature, and the neural network is then retrained using the new equilibrated configurations as the starting point for the next step. \new{Since the neural network produces entirely new configurations, the annealing procedure includes moves that can, in principle, update all the spins. Hence, the term \textit{global}, to distinguish the updates from the single-variable local updates used in algorithms such as the classical Metropolis Monte Carlo.}
The original implementation of GA did not include the local steps and relied only on global moves (akin to an importance sampling) to equilibrate the newly generated configurations. While this procedure is exact if the thermalization is carried out for long enough times, in practice the time taken to achieve equilibrium can be very long.

The work of Gabrié \textit{et al.} \cite{gabrie2022adaptive} has highlighted the importance of alternating global, NN-assisted moves and standard local Monte Carlo moves. In \cite{gabrie2022adaptive}, the authors study the convergence properties of a NN-assisted Monte Carlo procedure to a target distribution $\rho^*$. They provide numerical examples and analytical computations highlighting the faster convergence to $\rho^*$ when local moves are alternated to the NN-based global ones. However, at variance with the standard GA procedure, in Ref.~\cite{gabrie2022adaptive} no annealing in temperature is performed.
The additional empirical evidence presented in subsequent studies~\cite{hunt2024accelerating, mehdi2024enhanced} underscores the need for physical intuition and a clear theoretical framework, both of which remain elusive in the context of the GA procedure.

\subsection{\new{Summary of results and outline of the work}}

In this work, \new{we support the claim that local moves are necessary also in the context of GA}  by a complete theoretical analysis of the training and application of a shallow MADE (Masked Autoencoder for Distribution Estimation) neural network to assist sampling in the Curie-Weiss model. \new{In particular, we focus on a single step of the Global Annealing procedure starting from the critical temperature of the model. This choice is motivated by the fact that in the Curie-Weiss model, this is the hardest temperature to work at, because the free energy at zero magnetization becomes locally flat (with vanishing first and second derivatives), which slows down equilibration. Immediately below, a pitchfork bifurcation happens, and two new states (the states with positive and negative magnetizations) appear. Being able to bypass the critical slowing down and correctly identify and follow the new states that appear upon lowering the temperature is generally the bottleneck in any annealing procedure; therefore, testing the Global Annealing at the critical temperature provides a good indication of the efficiency of the whole procedure.}

\new{More generally, we believe that this is an important first step towards understanding more complex systems, such as spin glasses, random optimization problems, and neural networks, in which flat directions in the (free) energy landscape play a very important role in the dynamical slowing down, especially in the low temperature region where a large number of states coexist \cite{cugliandolo1993analytical,ricci2010being,montanari2004cooling,folena2020rethinking,baldassi2021unveiling,ly2025optimization}.}

Our main results are the following:
\begin{itemize}
    \item We give a full analytical description of the \new{shallow} MADE architecture \new{(Sec. \ref{sec}: Eq. \ref{eq:optimal_weights_simplified} and Fig. \ref{fig:Jl_diffbeta})} and of its training dynamics \new{(Sec. \ref{sec:training_dynamics}: Eq. \ref{eq:training_solution} and Fig. \ref{fig:comparison_approximation})} for the Curie-Weiss model, both for finite system sizes and in the thermodynamic limit;
    \item We show that a phenomenon akin to critical slowing down happens in learning at the critical temperature of the model \new{(Sec. \ref{sec:training_dynamics}: Eq. \ref{eq:learning_rate}), which once again highlights the special role played by this temperature as the bottleneck of the annealing};
    \item \new{We study one step of the Global Annealing procedure when lowering the temperature starting at the critical point. We show \new{(Sec. \ref{sec:perfectly_trained}: Fig. \ref{fig:FPT_diffB})} that a perfectly trained network does not require local moves as long as the temperature step is small enough (i.e. scaling as the inverse of the square root of the system size). Instead, local moves are beneficial once a finite training time is taken into account \new{(Sec. \ref{sec:non_perfectly_trained}: Fig. \ref{fig:ratios_trainingtime})};}
    \item We characterize the effectiveness of the Global Annealing procedure as compared to a standard local Metropolis Monte Carlo, in terms of first passage times in magnetization space \new{(Sec. \ref{sec:comparison_vanilla}: Fig. \ref{fig:com_vanilla})}.
\end{itemize}

In particular, our work extends some of the results in Ref.~\cite{ciarella2023machine}, in which the shallow MADE architecture was only studied in the $N \to \infty$ limit. Although an exact architecture for sampling according to the GB distribution can be constructed~\cite{biazzo2023autoregressive}, we focus here on a shallow MADE because it can be treated analytically also in the training \new{process. Since training plays a crucial role in practical applications involving complex systems with unknown exact parameters, studying the shallow MADE architecture allows us to better control and analyze a setting that closely resembles real-world scenarios.}

This paper is organized as follows. In Sec. \ref{sec:theoretical_background} we give the background for our work. In particular, we introduce the Curie-Weiss model (\ref{sec:curie_weiss}) and the algorithms we study, local Metropolis Monte Carlo (\ref{sec:localMMC}) and NN-assisted Monte Carlo (\ref{sec:sequential_tempering}) with a shallow MADE architecture (\ref{sec:made_description}). In Sec.~\ref{sec:methods} we present a theoretical analysis of the training and of the NN-assisted Monte Carlo. In particular, we study the optimal model and the training dynamics (\ref{sec:MADE}); moreover, we compare the performance of the algorithms in terms of first-passage times in magnetization space (\ref{sec:first_passage_times_analysis}). In Sec.~\ref{sec:results} we apply these methods to compare different sampling procedures. We first consider Global Annealing with a fully trained machine, with and without local MC steps (\ref{sec:perfectly_trained}); then, we study how the scenario changes when one takes into account a finite training time (\ref{sec:non_perfectly_trained}); additionally we compare Global Annealing with vanilla Metropolis Monte Carlo (\ref{sec:comparison_vanilla}). Finally, in Sec.~\ref{sec:conclusions} we draw our conclusions and highlight some possible future developments.

\section{Background}\label{sec:theoretical_background}
\subsection{The Curie-Weiss model}\label{sec:curie_weiss}

In this paper, we consider the simplest model of ferromagnetic phase transitions, the Curie-Weiss (CW) model. In the CW model, the state of the system is given by a set of $N$ Ising variables, $\boldsymbol{\sigma}$, $\sigma_i = \pm 1$, $i = 1, \dots, N$. The Hamiltonian reads:
\begin{equation}\label{eq:HCW}
    \mathcal{H}(\bs) = -\frac{N}{2}m(\bs)^2 \ ,
\end{equation}
where $m(\bs) = \frac{1}{N}\sum_i \sigma_i$ is the (intensive) magnetization of the system. It is well known that this model undergoes a phase transition in the thermodynamic limit at a critical temperature $T_c = 1$, passing from a disordered paramagnetic phase at $T > 1$ to an ordered ferromagnetic phase at $T < 1$ \new{(see Chapter 2 of \cite{friedli2018statistical} for a pedagogical introduction)}. In the ferromagnetic phase, the model develops a non-trivial spontaneous equilibrium magnetization, which is given by the non-zero solution $m^*$ of the equation
\begin{equation}\label{eq:m_CW}
    m^* = \tanh(\beta m^*) \ .
\end{equation}
The model can be simulated using one of many different algorithms. We focus on classical local Metropolis Monte Carlo,
a standard Swiss-knife algorithm for the simulation of statistical physics systems, 
and on a NN-assisted Monte Carlo procedure, Global Annealing. In particular, we want to study which algorithm is faster in equilibrating the system at $T < 1$ starting from $T = T_c = 1$, as measured using as a proxy the time needed to first reach the equilibrium magnetization that solves Eq.~\eqref{eq:m_CW}.
In the next section, we describe in detail the algorithms that we considered in our analysis.

\subsection{Algorithms}

\subsubsection{Standard local Metropolis Monte Carlo}\label{sec:localMMC}

In the standard local Metropolis Monte Carlo (LMMC) \cite{newman1999monte} algorithm, one performs a series of local, single-spin-flip moves. A single Monte Carlo `step' consists of the following operations, starting from a configuration $\bs(t) = \bs$ at time $t$:
\begin{enumerate}
    \item propose a new configuration by flipping the spin of a randomly chosen site, $\sigma_i \rightarrow -\sigma_i$;
    \item calculate the energy difference, $\Delta E = -2\left[\frac{1}{N}-\sigma_i m(\bs) \right]$, between the new configuration and the current configuration;
    \item accept the move, i.e. set $\bs(t+1)=\bs'$ where $\bs'$ is obtained from $\bs$ by flipping $\s_i$,     
    with probability:
    \begin{equation*}
\text{Acc}\left[\boldsymbol{\sigma} \rightarrow \boldsymbol{\sigma}'\right] = \min\left[1, e^{-\beta \Delta E}\right] \ .
    \end{equation*}
    Otherwise, reject the move and set $\bs(t+1)=\bs$.    
\end{enumerate}
$N$ such steps are commonly referred to as a Monte Carlo Sweep (MCS). It is easy to see that the computational complexity of a MCS is $\mathcal{O}(N)$.

Although LMMC frequently serves as a building block for more advanced and powerful algorithms, such as Parallel Tempering~\cite{hukushima1996exchange} and Population Annealing~\cite{hukushima2003population,barash2017gpu,martinez2025problem}, it may encounter limitations when employed in isolation.
For instance, it can remain trapped for long times in (local) minima of the free energy landscape, thus failing to sample effectively the whole space of configurations.\footnote{For instance, in the CW model below $T_c$, LMMC can remain stuck in the state with positive (negative) magnetization. Escaping the state and reaching the state with negative (positive) magnetization requires a time growing exponentially with the size of the system $N$.}

\subsubsection{The Global Annealing procedure}\label{sec:sequential_tempering}

The main disadvantage of the LMMC algorithm described in the previous section is that it can only perform local moves. A recently proposed solution is to use a generative NN to propose global moves. The idea is to start from a configuration $\bs$ and use a generative NN to generate a whole new configuration $\bs'$ of the system, which is then accepted with an acceptance ratio $\mathrm{Acc}$ chosen in order to satisfy detailed balance:
\begin{equation}\label{eq:prob}
\begin{split}
    \text{Acc}\left[\boldsymbol{\sigma} \rightarrow \boldsymbol{\sigma}'\right] &= \min\left[1, \frac{P_\text{GB}(\boldsymbol{\sigma}') \times P_{\text{NN}}(\boldsymbol{\sigma})}{P_\text{GB}(\boldsymbol{\sigma}) \times P_{\text{NN}}(\boldsymbol{\sigma}')}\right], 
    \end{split}
\end{equation}
where $P_{\text{NN}}(\boldsymbol{\sigma})$ is the probability that the NN generates the configuration $\boldsymbol{\sigma}$.
Note that in this case the whole configuration is updated in a single operation, which then corresponds roughly to a Monte Carlo Sweep.
While this strategy is essentially equivalent to an importance sampling of $P_\text{GB}(\boldsymbol{\sigma})$ using $P_{\text{NN}}(\boldsymbol{\sigma})$ as a generator, the formulation in terms of a stochastic process with an acceptance ratio allows one to combine the global moves with local ones, as we will do in the following.

The success of this strategy relies on the ability of the NN to generate configurations close to the ones at equilibrium, so that the full configuration space is well sampled and the acceptance rate is not too low. It has been shown \cite{biazzo2023autoregressive, del2024nearest} that it is possible to design networks that are powerful enough to sample according to the GB distribution. The question then becomes whether one is able to train such networks in practice. While variational procedures that do not require sampling have been used \cite{wu2019solving}, we focus here on methods that use previously generated equilibrium configurations to train the model, because variational methods have been shown to be prone to mode collapse in complex problems~\cite{ciarella2023machine}.

In these approaches, the NN is first trained at a temperature $\beta$ using a set of $M$ equilibrium configurations. The training can be carried out, for instance, by maximizing the model likelihood of the available $M$ equilibrium configurations.
Then, the NN is used to generate a new series of configurations, which are then used as proposal moves for a global Monte Carlo. The training set is updated using the acceptance rate in Eq.~\eqref{eq:prob} and the model is trained again. 
Unfortunately, obtaining the $M$ configurations used to train the model can be complicated, especially if one is interested in the low-temperature, hard-to-sample regime of a model.
And of course, if we are already able to obtain equilibrium configurations, there is no interest in developing new sampling methods.
A solution to both issues is to use the Global Annealing (GA) procedure, an annealing procedure that uses the self-consistently trained NN in order to generate configurations at lower and lower temperatures.

In GA, the NN is first trained using $M$ configurations at a high temperature, at which it is easy to sample configurations at equilibrium (for example via LMMC). Then, the NN is used at inverse temperature $\beta' = \beta+\Delta \beta > \beta$ to propose a new set of $M$ configurations at $\beta'$ by performing $\theta_\mathrm{global}$ global moves. This new set of configurations is then used to train a new NN (or retrain the previous one). The whole procedure can then be repeated until the desired temperature is reached. The general scheme of GA is summarized in Algorithm~\ref{alg:ST}. 
Additionally, $\theta_\mathrm{local}$ LMMC steps can be alternated to the global moves proposed by the NN (steps 9-11 of Alg.~\ref{alg:ST}) but not all implementations include them \cite{mcnaughton2020boosting, ciarella2023machine}. Understanding the importance of performing local moves is one of the goals of this paper.


\begin{algorithm}
\caption{Global Annealing}
\label{alg:ST}
\begin{algorithmic}[1]
    \State \textbf{Input:} Initial inverse temperature $\beta_\text{start}$, final inverse temperature $\beta_\text{end}$, temperature step $\Delta\beta$, number of configurations~$M$, number of global steps per temperature $\theta_\text{global}$, number of local steps per global step $\theta_\text{local}$.
    \State \textbf{Initialize:} A set of $M$ equilibrium configurations at $\beta_\text{start}$ (sampled e.g. using standard Metropolis MC)
    \While{$\beta < \beta_\text{end}$}
        \State Train a neural network (NN) using the set of $M$ configurations
        \State Lower the temperature: $\beta \gets \beta + \Delta \beta$
        \For{$m$ in $1,\dots,M$}
        \State Choose the $m$-th configuration from the set as the initial state
        \For{$t$ in $1, \dots, \theta_\text{global}$}
            \State Propose a new configuration using the NN
            \State Accept or reject the configuration with probability \eqref{eq:prob} at the new temperature $T=1/\beta$
            \For{$t$ in $1, \dots, \theta_\text{local}$}
            \State Perform a LMMC step
        \EndFor
        \EndFor
        \EndFor
    \EndWhile
\end{algorithmic}
\end{algorithm}

The GA scheme is quite general. The specific implementation then requires the choice of a generative NN. In this paper, we consider a shallow MADE, as described in the next section.

\subsection{The MADE architecture}
\label{sec:made_description}

We consider as our architecture of choice the shallow MADE (Masked Autoencoder for Distribution Estimation, \cite{germain2015made}) considered in Ref.~\cite{ciarella2023machine}.

The MADE is an autoregressive model, in which the probability of a configuration is represented as a sequence of conditional probabilities:
\begin{equation}
    P_\text{NN}(\boldsymbol{\sigma}) = P(\sigma_1) P(\sigma_2 \mid \sigma_1) P(\sigma_3 \mid \sigma_1, \sigma_2) \cdots P(\sigma_N \mid \sigma_{1}, \cdots, \sigma_{N-1}) = \prod_{i=1}^{N} P(\sigma_i \mid \boldsymbol{\sigma}_{<i}),
\end{equation}
and the $P(\sigma_i \mid \boldsymbol{\sigma}_{<i})$ are written in terms of a set of parameters that defines the model.
This formalization allows us not only to compute the probability of a given configuration $\boldsymbol{\sigma}$, but also to generate a new one from scratch in polynomial time using \textit{ancestral sampling}. In ancestral sampling, first the spin $\sigma_1$ is generated using $P(\sigma_1)$, then $\sigma_2$ is generated using $\sigma_1$ according to $P(\sigma_2 \mid \sigma_1)$, then $\sigma_3$ is generated using $\sigma_2$ and $\sigma_3$ according to $P(\sigma_3 \mid \sigma_1, \sigma_2)$ and so on.

Specifically, the shallow MADE we are considering parametrizes the probability as $P(\s_1)=1/2$ and, for $i>1$,
\begin{equation}
    P(\sigma_i | \bs_{<i}) = \frac{\exp\left( \sum_{j=1}^{i-1} J_{i\new{j}} \sigma_i \sigma_j \right)}{2 \cosh\left( \sum_{j=1}^{i-1} J_{ij} \sigma_j \right)}
\end{equation}
\new{where $J_{ij}$ are learnable weights. We make a further simplification, motivated by permutation invariance of the original model, that the weights are shared across each row, so that $J_{ij} = J_{i}$ and}
\begin{equation}
    \new{P(\sigma_i | \bs_{<i}) = \frac{\exp\left( \sum_{j=1}^{i-1} J_i \sigma_i \sigma_j \right)}{2 \cosh\left( \sum_{j=1}^{i-1} J_i \sigma_j \right)} = \frac{\exp(J_i \sigma_i M_{<i})}{2 \cosh(J_i M_{<i})},}
\end{equation}
where $M_{<i}(\boldsymbol{\sigma}_{<i}) = \sum_{j=1}^{i-1} \sigma_i$. This architecture corresponds to a NN with a \new{single dense autoregressive} layer with shared \new{per-row} weights \new{(i.e. all the weights for each row are the same, so that, in practice, for each conditional probability one sums the previous input variables and then multiplies the sum by a single learnable weight)}, followed by a softmax activation function, and is fully specified by a set of $N-1$ weights $\underline{J} = (J_2,\cdots,J_N)$, \new{that is, a number of parameters linear in the system size}. \new{This setup differs from what is usually applied in deep learning, in which networks have a large number of parameters, a feature that seem to be connected to their ability to be trained to reach good generalization even in situations in which the loss is strongly nonconvex (the interested reader can refer to \cite{baldassi2021unveiling, ly2025optimization} and references therein). While the shallow MADE architecture cannot leverage the properties of deep NNs, the problem at hand is simple enough that this architecture is adequate~\cite{ciarella2023machine}.} \new{Moreover, as explained in Sec. \ref{sec:MADE}, the best set of parameters of the model can be found by maximizing the likelihood of the training data and for this shallow MADE applied to the Curie-Weiss model this optimization can be written down explicitly, as we do in the next section. This allows us to get insight that can be applied to more complicated architectures, as well to more complex physical systems, whose study is left to future work.}

\section{Methods}\label{sec:methods}
\subsection{Analysis of the MADE architecture}\label{sec:MADE}
\subsubsection{Optimal values of the weights}\label{sec}

\begin{figure}[t]
    \centering
    \includegraphics[width=1\linewidth]{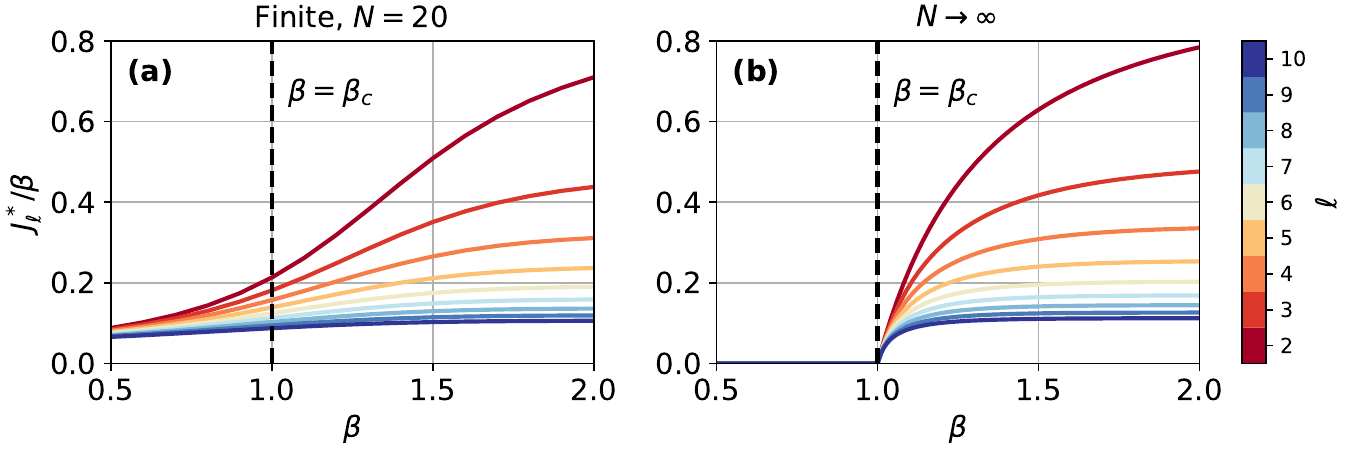}
    \caption{Behavior of the optimal couplings $J_\ell^*/\ell$ as a function of $\beta$ for $\ell \leq 10$. (a) Finite $N$, ($N = 20$), obtained solving Eq.~\eqref{eq:optimal_weights}. (b) Infinite $N$, obtained solving Eq.~\ref{eq:optimal_weights_Ninf}.}
    \label{fig:Jl_diffbeta}
\end{figure}

In order to analyze the MADE architecture described in Sec.~\ref{sec:made_description}, let us start by introducing the cross entropy $S_c$ between $P_\text{GB}$ and $P_\text{MADE}$ for a set of weights $\underline{J}$, 
\begin{equation}\label{eq:KL_loss}
\begin{split}
    S_c(\underline{J}) &= - \sum_{\{\boldsymbol{\sigma}\}} P_\text{GB}(\boldsymbol{\sigma}) \log P_\text{MADE}(\boldsymbol{\sigma}) 
= - \sum_{\{\boldsymbol{\sigma}\}} P_\text{GB}(\boldsymbol{\sigma}) 
\log\left[ \frac12 \prod_{i=2}^N P(\sigma_i | \boldsymbol{\sigma}_{<i})\right] \\
&= N \log 2 - \sum_{i=2}^N
\sum_{\{\boldsymbol{\sigma}\}} P_\text{GB}(\boldsymbol{\sigma})
\left\{ J_i \sigma_i M_{<i} - \log[  \cosh(J_i M_{<i})]   \right\} \ .
\end{split}
\end{equation}
The sum over $\boldsymbol{\sigma}$ runs over all $2^N$ possible configurations of the system. 
Minimizing the cross entropy with respect to the $\ell$-th coupling $J_\ell$ yields:
\begin{equation}\label{eq:zero_gradient}
    \sum_{\{\boldsymbol{\sigma}\}} P_\text{GB}(\boldsymbol{\sigma}) \, M_{<\ell} \sigma_\ell = \sum_{\{\boldsymbol{\sigma}\}} P_\text{GB}(\boldsymbol{\sigma}) \, M_{<\ell} \tanh(J_\ell M_{<\ell}) \ .
\end{equation}
In the CW model, the sum over the $2^N$ spin configurations can be reduced to a polynomial sum by rewriting
the Gibbs-Boltzmann distribution $P_\text{GB}$ in Eq.~\eqref{eq:GB_distribution} with the CW Hamiltonian in Eq.~\eqref{eq:HCW} via a Hubbard-Stratonovich transformation as:
\begin{equation}\label{eq:PGB_HS}
    P_\text{GB}(\boldsymbol{\sigma}) = \frac{1}{\hat{\mathcal{Z}}(\beta)}\int dh \, e^{-\frac{N h^2}{2\beta}} e^{h \sum_i \sigma_i} \ ,
\end{equation}
where $\hat{\mathcal{Z}}(\beta) = \mathcal{Z}(\beta)\sqrt{\frac{2 \pi \beta}{N}}$ is the new normalizing constant. Inserting this expression in Eq.~\eqref{eq:zero_gradient} one finds:\footnote{\new{The result follows from explicitly computing the sums over the spins $\geq \ell$ while rewriting the sums over spins $< \ell$ in terms of sums over the possible magnetizations.}}
\begin{equation}\label{eq:optimal_weights}
    \int dh \, e^{-\frac{N h^2}{2 \beta}} \sinh(h) \cosh^{N-\ell}(h) \, \Sigma_1(h)
    = \int dh \, e^{-\frac{N h^2}{2 \beta}} \cosh^{N-\ell+1}(h) \, \Sigma_2(h, J_\ell) \ ,
\end{equation}
where
\begin{equation}
    \Sigma_1(h) = \sum_{\substack{M = -(\ell-1) \\ M +=2}}^{\ell-1} \binom{\ell-1}{\frac{\ell - 1 - M}{2}} e^{h M } M
\qquad
\text{ and }
\qquad
    \Sigma_2(h, J_\ell) = \sum_{\substack{M = -(\ell-1) \\ M +=2}}^{\ell-1} \binom{\ell-1}{\frac{\ell - 1 - M}{2}} e^{h M} M \tanh(J_\ell M) \ ,
\end{equation}
where the magnetization $M$ increases in steps of 2 in the sums.
Note that with some manipulations, Eq.~\eqref{eq:optimal_weights}
can be simplified to\footnote{\new{The simplification comes from the fact that the sum that defines $\Sigma_1$ can be computed explicitly and yields:}
\begin{equation*}
    \new{\Sigma_1(h) = \sum_{\substack{M = -(\ell-1) \\ M +=2}}^{\ell-1} \binom{\ell-1}{\frac{\ell - 1 - M}{2}} e^{h M } M = \sum_{k=0}^{\ell-1}\binom{\ell-1}{k}e^{h(\ell-1-2k)}(\ell-1-2k) = 2^{\ell-1}(\ell-1)\cosh^{\ell-2}(h)\sinh(h).}
\end{equation*}
}
\begin{equation}\label{eq:optimal_weights_simplified}
    2^{\ell-1}(\ell-1)\int dh \, e^{-\frac{N h^2}{2 \beta}} \sinh^2(h) \cosh^{N-2}(h)
    = \int dh \, e^{-\frac{N h^2}{2 \beta}} \cosh^{N-\ell+1}(h) \, \Sigma_2(h, J_\ell) \ .
\end{equation}
The latter equations involve a single integral and a sum over a linear number of terms in $N$, and can thus be solved numerically to find the optimal values of the weights $J_\ell$. An example of the behavior of $J_\ell$ as a function of $\beta$ is shown in Fig. \ref{fig:Jl_diffbeta}a.

\subsubsection{Thermodynamic limit}\label{sec:thermodynamic}

In the $N \to \infty$ limit the integrals over $h$ can be evaluated using the Laplace method, and Eq.~\eqref{eq:optimal_weights} reduces to
\begin{equation}\label{eq:optimal_weights_Ninf}
    \Sigma_2(h^*, J_\ell) = \Sigma_1(h^*) \tanh(h^*)\ ,
\end{equation}
where $h^*$ is the solution of the saddle-point equation
\begin{equation}\label{eq:hstar}
    h^* = \beta \tanh(h^*) \ .
\end{equation}
Notice that we can consider just the positive solution for $h^*$, since taking into account the negative one simply yields additional factors of two on both sides of the equations.
These results match the equations derived in Ref.~\cite{ciarella2023machine} for the $N \to \infty$ limit. 
An example of the behavior of the weights in the infinite-$N$ limit is shown in Fig.~\ref{fig:Jl_diffbeta}b.

\begin{figure}
    \centering
    \includegraphics[width=1\linewidth]{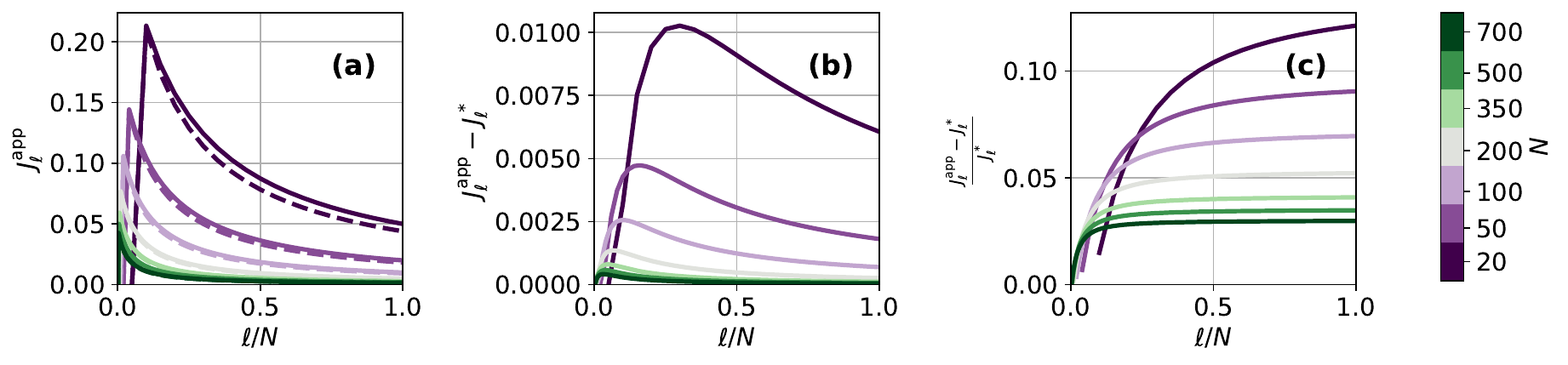}
    \caption{Comparison \new{at $\beta = \beta_c = 1$} of the approximated couplings $J^\text{app}_\ell$ as found by solving Eq.~\eqref{eq:small_Jl} with the exact ones. (a) $J^\text{app}_\ell$ (dashed) compared with the exact $J^*_\ell$ (full). (b) Absolute error $J^*_\ell-J^\text{app}_\ell$. 
    (c) Relative error $(J^*_\ell-J^\text{app}_\ell)/J^*_\ell$.}
    \label{fig:approximation}
\end{figure}

For $\beta \leq \beta_c = 1$,
Eq.~\eqref{eq:hstar} only admits the $h^* = 0$ solution. As a consequence, $J_\ell = 0$ $\forall \ell$, i.e. for $N \to \infty$ all the weights vanish. We can then try a small $J_\ell$ approximation of Eq.~\eqref{eq:optimal_weights} at finite $N$. At first order, this yields the equation:
\begin{equation}\label{eq:small_Jl}
    J^\text{app}_\ell = \frac{\int_{-\infty}^{\infty} \exp\left(-\frac{N}{2\beta} h^2\right) \cosh^{N-2}(h) \sinh^2(h)  \, dh}{\int_{-\infty}^{\infty} \exp\left(-\frac{N}{2\beta} h^2\right) \cosh^{N-2} (h)\left( (l-2)\sinh(h)^2 + \cosh(h)^2 \right ) \, dh} .
\end{equation}
A comparison between the approximated weights and the exact ones is shown in Fig.~\ref{fig:approximation}.
In the following section, we study how the weights of the model are learned during training.

\subsubsection{Training dynamics}\label{sec:training_dynamics}

We consider a gradient descent (GD) training dynamics, taking as a loss the cross entropy defined in Eq.~\eqref{eq:KL_loss}. Then, the update rule of the parameters can be written as:
\begin{equation}\label{eq:update_weights}
    J_\ell^{(t+1)} = J_\ell^{(t)}-\eta_\ell\nabla_\ell S_c(\underline{J}^{(t)}) \ ,
\end{equation}
where $\eta_\ell$ is the learning rate for the $\ell-$th weight and the gradient is:
\begin{equation}
\nabla_\ell S_c(\underline{J}) =  \frac{2^{N-\ell}}{\mathcal{Z}(\beta)}\sqrt{\frac{2N}{\pi \beta}} 
\int_{-\infty}^{\infty} dh \, e^{-\frac{N h^2}{2 \beta}} \cosh(h)^{N - \ell}
\sum_{\substack{M= -(\ell - 1) \\ M += 2}}^{\ell - 1} 
\binom{\ell - 1}{\frac{\ell - 1 - M}{2}} e^{h M} M 
\left [ \cosh(h) \tanh(J_\ell M)-\sinh(h) \right ] \ .
\end{equation}
Note that because the cross entropy is a sum of terms, each involving a single weight, the gradient $\nabla_\ell S_c(\underline{J})$ depends only
on $J_\ell$ and the gradient descent dynamics of different weights are decoupled.

\begin{figure}[t]
    \centering
    \includegraphics[width=0.6\linewidth]{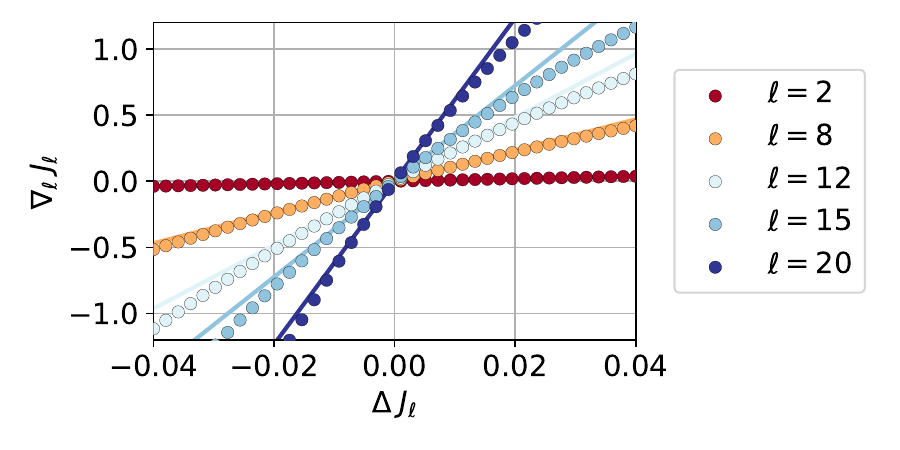}
    \caption{Comparison between the gradients obtained by linearizing around the optimal solution $J_\ell^*$ (full lines) and the gradients computed using \texttt{pytorch} backpropagation on a large dataset (data points) as a function of the distance from the optimal couplings, $\Delta J_\ell = J_\ell-J^*_\ell$. Details: $N = 20$ spins, $\beta = 1$, the dataset is made of $5 \cdot 10^6$ configurations obtained by starting at infinite temperature and then performing 30 MCS at $\beta = 1$.}
    \label{fig:gradients}
\end{figure}

In the continuous time limit (gradient flow), Eq.~\eqref{eq:update_weights} reads:
\begin{equation}\label{eq_GD_continuous}
    \dot{J}_\ell = -\eta_\ell \nabla_\ell  S_c(\underline{J}) \ .
\end{equation}
The gradient can be linearized around the solution $J_\ell^*$ of Eq.~\eqref{eq:optimal_weights}, yielding:
\begin{equation}
    \dot{J}_\ell = -\eta_\ell \Delta J_\ell\mathrm{H}_\ell(J^*_\ell)  \ ,
\end{equation}
where $\Delta J_\ell = J_\ell-J^*_\ell$ is the difference with respect to the optimal couplings and $\mathrm{H}_\ell$ is the second derivative of the cross
entropy, given by:
\begin{equation}
\mathrm{H}_\ell(J_\ell) = \frac{2^{N - \ell}}{\mathcal{Z}(\beta)} 
\int_{-\infty}^{\infty} dh \, 
e^{-\frac{N h^2}{2 \beta}} 
\sqrt{\frac{2 N}{\pi \beta}} \, 
\cosh(h)^{N - \ell + 1}
\sum_{\substack{M = -(\ell - 1) \\ M += 2}}^{\ell - 1} 
\binom{\ell - 1}{\frac{\ell - 1 - M}{2}} 
e^{h M} \, M^2 \, \text{sech}^2(J_\ell M) \ .
\end{equation}
A comparison between the linearized gradient and the true gradient is shown in Fig.~\ref{fig:gradients}, highlighting the very good agreement between the two. 

\begin{figure}[t]
    \centering
    \includegraphics[width=0.6\linewidth]{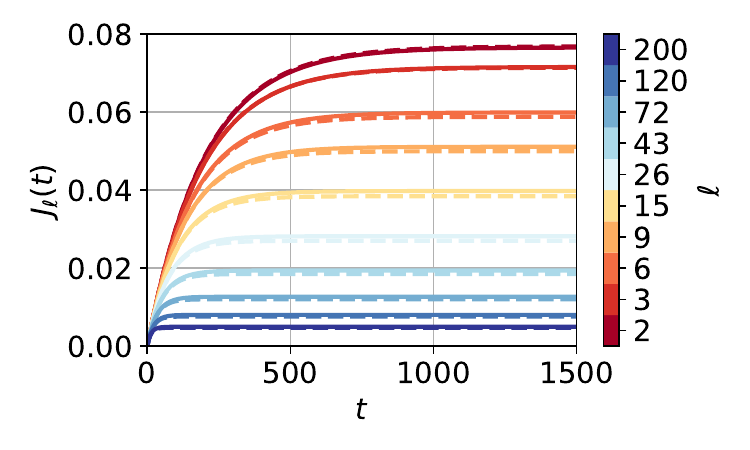}
    \caption{Comparison between the training of the weights obtained by the approximation in Eq.~\eqref{eq:training_solution} (full lines) and the training performed numerically using \texttt{pytorch} over a large dataset. Details: $N = 200$ spins, $\beta = 1$, the dataset is made of $5 \cdot 10^6$ equilibrium configurations, learning rate $\eta_\ell = 1/[N(\ell-1)]$.}
    \label{fig:comparison_approximation}
\end{figure}

Unfortunately, this approach is not easily tractable analytically. Instead, for $\beta \leq \beta_c=1$, we can consider the small $J^*_\ell$ approximation and linearize around zero. By linearizing Eq.~\eqref{eq_GD_continuous} (in the small $J^*_\ell$ approximation), one finds:
\begin{equation}
    \dot{J}_\ell = -\eta_\ell \left [ \< M_{<\ell}\sigma_\ell \>- \mathrm{H}_\ell(0)J_\ell \right] = -\eta \left [ (\ell-1) c_N- \mathrm{H}_\ell(0)J_\ell \right] \ ,
\end{equation}
where $\mathrm{H}_\ell(0)$ has now the simple form:
\begin{equation}
    \mathrm{H}_\ell(0) = (\ell-1) \left [ 1+(\ell-2)c_N\right],
\end{equation}
and $c_N$ is the two-spin correlation function $c_N = \<s_i s_j\>$, $i \neq j$ between any two different spins,
\begin{equation}
    c_N = \<s_i s_j\> = \frac{\displaystyle \int_{-\infty}^{\infty} e^{-\frac{N h^2}{2\beta}} \sinh^2(h) \cosh^{N-2}(h) \, dh}
     {\displaystyle \int_{-\infty}^{\infty} e^{-\frac{N h^2}{2\beta}} \cosh^N(h) \, dh} \ ,
\end{equation}
which decays as $1/N$ for $T < 1$ and as $1/\sqrt{N}$ at $T = 1$.
The latter approximation allows for the training dynamics to be solved explicitly:
\begin{equation}\label{eq:training_solution}
J_\ell(t) = c_N \tau_\ell(\ell-1)\left [1-e^{-\frac{\eta_\ell t}{\tau_\ell}} \right] \ ,
\end{equation}
where the characteristic time $\tau_\ell$ is simply the inverse of the Hessian, $\tau_\ell = 1/\mathrm{H}_\ell(0)$, and $\eta_\ell$ is the learning rate. Notice that the prefactor $J^\text{app}_\ell = c_N \tau_\ell(\ell-1) = c_N/(1+(\ell-2)c_N)$ corresponds exactly to the small $J_\ell$ approximation derived in Eq.~\eqref{eq:small_Jl}.
While this argument can probably be made more rigorous, e.g. using the Polyak–Łojasiewicz inequality, we instead verify the correctness of this assumption numerically in Fig. \ref{fig:comparison_approximation}, finding an excellent agreement between the exact and approximate solutions.

\begin{figure}[t]
    \centering
    \includegraphics[width=1\linewidth]{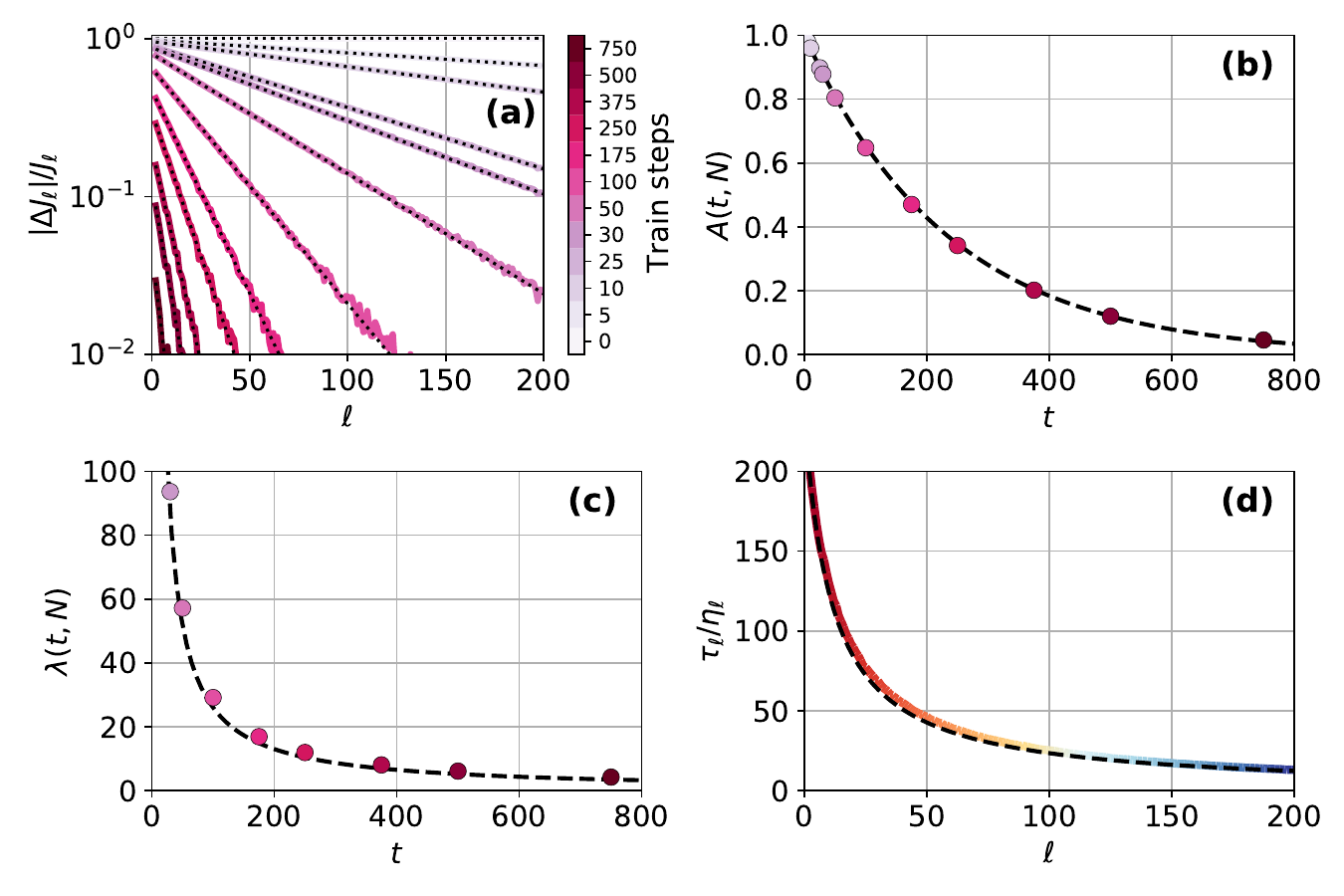}
    \caption{Comparison between theory and numerical results for different quantities. The data used come from the same training of Fig.~\ref{fig:comparison_approximation}.
    (a) Relative error $|\Delta J_{\ell}|/J_{\ell}$ plotted as a function of $\ell$, together with an exponential fit  to the form $Ae^{-\frac{\ell}{\lambda}}$ (dashed black lines). (b,c) The values of the fitted parameters $A$ and $\lambda$ (data points) are compared with those derived from the theory (dashed black lines). (d) The effective timescale $\hat\tau_\ell =\tau_\ell/\eta_\ell$, obtained fitting the relative error as $\frac{|\Delta J_\ell|}{J_\ell} = e^{-\frac{t}{\hat\tau_\ell}}$, is compared to the prediction from the theory.}
    \label{fig:different_quantities}
\end{figure}

Equation \eqref{eq:training_solution} allows to predict several important trends:
\begin{itemize}
    \item the relative error vanishes exponentially in time, as $\frac{|\Delta J_\ell|}{J_\ell} = e^{-\frac{\eta_\ell t}{\tau_\ell}}$;
    \item if $\eta_\ell = 1/[N(\ell-1)]$ then 
    $\frac{|\Delta J_\ell|}{J_\ell} = e^{ - \frac{1+(\ell-2)c_N}{N} t }=
    A(t, N)e^{-\frac{\ell }{\lambda(t,N)}}$, where $\log A(t, N) = \frac{t}{N}(2c_N-1)$ and $\lambda(t, N) = \frac{N}{t c_N}$; hence, the relative error also vanishes exponentially in $\ell$ at fixed time;
    \item again, if $\eta_\ell = 1/[N(\ell-1)]$, then the effective time scale is $\hat\tau_\ell = \frac{\tau_\ell}{\eta_\ell}  =\frac{N}{1+(\ell-2)c_N}$.
\end{itemize}
These predictions are verified in Fig. \ref{fig:different_quantities}.

Notice that Eq.~\eqref{eq:training_solution} requires to specify the learning rate $\eta_\ell$ for weight $\ell$. From a discretization of the gradient flow Eq.~\eqref{eq:training_solution}, noticing that in GD one performs a single discrete step at each time so that the minimum increment in $t$ is one, it follows that the learning rate must be taken as
\begin{equation}\label{eq:prescription_LR}
    \eta_\ell \sim \tau_\ell = 1/\mathrm{H}_\ell(0) \ .
\end{equation}
This result matches the known one for convex optimization problems~\cite{mehta2019high}. Indeed, this choice allows one to learn all the weights in a time $\mathcal{O}(1)$. However, this requires the knowledge of the Hessian, which is not usually known, and the choice in Eq.~\eqref{eq:prescription_LR} is therefore unrealistic. 

\begin{figure}
    \centering
    \includegraphics[width=0.5\linewidth]{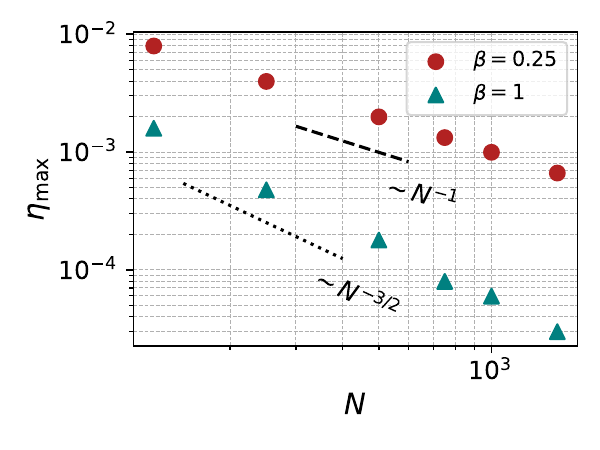}
    \caption{Maximum learning rate $\eta_{\max}$ as a function of system size $N$ at high temperature ($\beta = 0.25$) and at criticality ($\beta = 1$). The black segments highlight the scaling laws. The maximum learning rate is estimated by evaluating GD convergence as follows: we start from zero couplings, perform two numerical GD steps (using 50\,000 equilibrium configurations), and check whether the couplings remain positive (negative couplings after two GD steps indicate that GD does not converge). We then perform a bisection search to find the largest learning rate that still allows convergence.}
    \label{fig:maxeta_vs_N}
\end{figure}

In a more realistic implementation (at least for simple GD optimization), which we therefore consider in the following, one would use a single learning rate $\eta$ for all the weights. Then, the prescription for smooth convergence in Eq.~\eqref{eq:prescription_LR} becomes
\begin{equation}\label{eq:learning_rate}
    \eta \leq \tau_N \sim \frac{1}{c_NN^2} \ ,
\end{equation}
because the learning rate must be smaller than the smallest timescale in order for all the weights to be able to converge. At the critical temperature $T = T_c$, since $c_N \sim\frac{1}{\sqrt{N}}$ (see, for instance, Ref.~\cite[Eq.~23]{deger2020lee}), 
we need $\eta \sim N^{-\frac{3}{2}}$, which is the learning rate we will consider in the following. Notice that, since the slowest timescale (corresponding to $\ell = 2$) is of order 1, the effective timescale to learn all the weights goes as $N^\frac{3}{2}$. On the other hand, for $T \geq T_c$ (that is when we are no longer in the critical regime), $c_N \sim \frac{1}{N}$ and therefore the maximum learning rate that can be chosen goes as $\eta \sim \frac{1}{N}$, so that the timescale to learn all the weights is $\mathcal{O}(N)$ (see Fig. \ref{fig:maxeta_vs_N}). Therefore, at criticality, proper training requires an additional factor $\sqrt{N}$ in training time. This factor is exactly the same that appears due to critical slowing down (with exactly the same dynamical critical exponent) when performing Glauber dynamics or Metropolis-Hastings Monte Carlo \cite{bierkens2017piecewise}. So, in this setting, the hardness of sampling at criticality is instead transferred to the training.\footnote{Notice that we can rewrite the scaling of the learning rate around the critical temperature using a scaling function. In particular, we can write $\eta = N^{-3/2}f(\sqrt{N}\Delta T)$, where $\Delta T = T-T_c$, $f(0) = 1$ and $f(x) \sim x$ for $x \to \infty$. Therefore, the effect of criticality can be felt inside a window of size $1/\sqrt{N}$. As we will see in Sec.~\ref{sec:perfectly_trained}, this is the natural scaling of the temperature steps in the annealing procedure, so one cannot avoid the critical region by simply taking a larger temperature step.} It would be interesting to verify whether this scenario is generically present for different models and architectures. This analysis is left for future work.

\subsection{Analysis of Global Annealing and first passage times}\label{sec:first_passage_times_analysis}

We are interested in the time taken by the system to generate a configuration of a given (absolute) magnetization. Therefore, we can consider the dynamics of the model in the space of (intensive) magnetizations, which can be described in terms of a simple one-dimensional Markov chain. Then, we consider the time required by the chain to first reach a magnetization equal or greater (\new{possibly} in modulus) than a target magnetization. At fixed $\beta$, we take as the target magnetization the equilibrium magnetization, i.e. the solution $m^*$ of Eq.~\eqref{eq:m_CW}.

\new{We start by noting that} the average first-transition times $\tau_{m \to m^*}$ for going from a magnetization $m$ to a magnetization $m^*$ can be obtained using the set of self-consistent equations \cite{hunter2018computation}:
\begin{equation}\label{eq:first_passage}
    \tau_{m \to m^*} = 1 + \sum_{\hat{m} \neq m^*} P(m \to \hat{m}) \tau_{\hat{m} \to m^*}, \; \; \; \; \; m \neq m^* \ .
\end{equation}
If we call $\boldsymbol{\tau}$ the vector of the average first passage times (excluding the first passage time $m^* \to m^*$) and $\mathbb{Q}$ the matrix of the transition probabilities with the row and column corresponding to $m^*$ removed, we can rewrite the system in matrix form as:
\begin{equation}\label{eq:vectimes}
    (\mathbb{I}-\mathbb{Q}) \boldsymbol{\tau} = \mathbf{1} \ ,
\end{equation}
where $\mathbb{I}$ is the identity matrix and $\mathbf{1}$ is the vector of all ones. In practice, since we are interested in reaching magnetizations \new{$m \geq m^*$}, Eq.~\eqref{eq:first_passage} reduces to 
\begin{equation}\label{eq:transition_times}
    \tau_{m \to m^*} = 1 + \sum_{\hat{m} < m^*} P(m \to \hat{m}) \tau_{\hat{m} \to m^*} \ , \; \; \; \; \; m \neq m^* \ ,
\end{equation}
and therefore the size of the matrix that needs to be effectively inverted is smaller. \new{We will use signed magnetizations for the comparison with standard LMMC in Sec. \ref{sec:comparison_vanilla}. In Sec. \ref{sec:perfectly_trained} and \ref{sec:non_perfectly_trained} we will use the absolute values instead. Equation \eqref{eq:transition_times} stays the same but in terms of absolute magnetizations $|m|$ instead of signed magnetizations $m$.}  \new{The  transition matrix in \eqref{eq:transition_times} is then farther reduced by looking at the space of absolute magnetizations and, therefore, at the transition times to reach $|m| \geq |m^*|$}.

Notice that, if the matrix $\mathbb{I}-\mathbb{Q}$ is tridiagonal (as in the case for local single-spin flip algorithms), Eq.~\eqref{eq:vectimes} can be solved in linear time, either by inverting the matrix \cite{usmani1994inversion, da2007eigenvalues}, or using Thomas' algorithm \cite{thomas1949elliptic, tian2024swpts}.

We point out that, in alternative to the described procedure, one could evaluate the equilibration time also by looking at the second largest eigenvalues of the probability transition matrix. The advantage of looking at first passage times, however, is that if $m$ is not too large, one can work with an effective matrix that is smaller than the original one, thus making the computation easier.

\begin{figure}[t]
    \centering
    \includegraphics[width=0.5\linewidth]{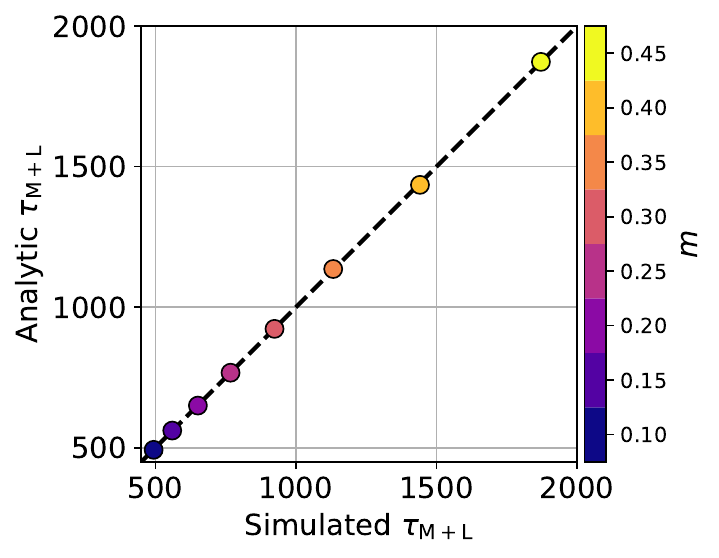}
    \caption{Comparison between the first passage times \new{it takes for the procedure that alternates a global MADE move and a local MCS} to reach magnetization $m$ starting from zero magnetization, $\tau_\mathrm{M+L}$ \new{(where M+L stands for MADE+LMMC)}. The values computed analytically are plotted versus those obtained numerically. Data are for $N = 200$ and $\beta = 1.1$. The averages are performed over $5 \cdot 10^4$ runs. Both times were multiplied by $2N$ to take into account the computational complexity of each move.}
    \label{fig:example_FPT}
\end{figure}

We consider three Monte Carlo schemes, in which:
\begin{itemize}
    \item only MADE global steps are performed;
    \item only LMMC steps are performed;
    \item a MADE step is followed by one or more LMMC sweeps.
\end{itemize}
An example of a comparison of the first passage times obtained by the \new{analytical} procedure described above and those obtained by performing a simulation in the case in which a MADE step is followed by one LMMC sweep is shown in Fig. \ref{fig:example_FPT}. \new{Just for this plot, we did not fix the target magnetization as the equilibrium one, but instead fixed $\beta=1.1$ and considered different values of $m$. We stress that it is possible to reach magnetizations larger than the equilibrium one since we are considering a finite-size system, so that the probability of reaching an arbitrarily large magnetization by a random fluctuation is always non-zero.}

The transition matrices obtained in the three cases are described in the following sections.

\subsubsection{Local Metropolis Monte Carlo}

We recall that for the single-spin-flip Metropolis MC one selects one spin at random and flips it with probability:
\begin{equation}\label{eq:metropolis}
    \text{Acc}\left[\boldsymbol{\sigma} \rightarrow \boldsymbol{\sigma}'\right] = \min\left[1, e^{-\beta \Delta E}\right] \ .
\end{equation}
Let us consider a configuration with magnetization $m$ and suppose one randomly selects a spin $\sigma$. The change in energy if the spin is flipped is:
\begin{equation}\label{eq:diff_energy}
    \Delta E = -2\left(\frac{1}{N}-\sigma m \right) \ .
\end{equation}
If $m = 0$, $\Delta E < 0$ and the move is always accepted. When $|m| > 0$ and one selects a spin at random, the probability of it having a sign opposite to $m$ is $\frac{1-|m|}{2}$ (and that of having the same sign is $\frac{1+|m|}{2}$). Using this and equations \eqref{eq:metropolis} and \eqref{eq:diff_energy}, we can write the transition probability of the Markov chain, keeping in mind that, since the dynamic is local, jumps in magnetization only occur between configurations separated by $\Delta m = 2/N$:
\begin{equation}
P(m \to m') =
\begin{cases}
P(m \to m - \frac{2}{N}) = \frac{1 + m}{2} \min[1, \exp\left( -2\beta \left( m - \frac{1}{N} \right) \right)], \\
P(m \to m - \frac{2}{N}) = \frac{1 - m}{2} \min[1, \exp\left( 2\beta \left( m + \frac{1}{N} \right) \right)], \\
P(m \to m) = 1-\frac{1 + m}{2} \min[1, \exp\left( -2\beta \left( m - \frac{1}{N} \right) \right)]-\frac{1 - m}{2} \min[1, \exp\left( 2\beta \left( m + \frac{1}{N} \right) \right)].
\end{cases}
\end{equation}
Analogously, the transition matrix in the space of absolute magnetizations $|m|$ is:
\begin{equation}\label{eq:MarkovChain}
P(|m| \to |m'|) = 
\begin{cases}
P(0 \to \frac{2}{N}) = 1, & \text{for } |m| = 0, \\
\begin{cases}
P(|m| \to |m| + \frac{2}{N}) = \frac{1 - |m|}{2}, \\
P(|m| \to |m| - \frac{2}{N}) = \frac{1 + |m|}{2} e^{-2 \beta (|m| - \frac{1}{N})}, \\
P(|m| \to |m|) = 1 - \left( \frac{1 - |m|}{2} + \frac{1 + |m|}{2} e^{-2 \beta (|m| - \frac{1}{N})} \right),
\end{cases}
& \text{for } |m| > 0, \\
P(|m| \to |m|') = 0, & |m-m'| \neq 0, \, 2/N.
\end{cases}
\end{equation}
Interestingly, taking the $N \to \infty$ limit and requiring the probabilities of increasing and decreasing the magnetization to be equal yields:
\begin{equation}
    \frac{1 + |m|}{2} e^{-2 \beta |m|} = \frac{1 - |m|}{2} 
    \qquad \Leftrightarrow
    \qquad 
    |m| = \tanh \beta |m| \ ,
\end{equation}
which is the correct equation for the equilibrium magnetization in the Curie-Weiss model, Eq.~\eqref{eq:m_CW}.

\subsubsection{MADE}

The transition matrix in the case of the MADE can be written as:
\begin{equation}\label{eq:MarkovChainGlobal}
P(|m| \to |m'|) = 
\begin{cases}
\frac{\Omega(|m'|)}{\Omega(|m|)} \min\left(1, \frac{p_\text{MADE}(|m'|)}{p_\text{MADE}(|m|)} e^{\frac{\beta N}{2}(|m'|^2 - |m|^2)}\right) p_\text{MADE}(|m'|)\ , & |m| \neq |m'| \ , \\
1 - \sum_{|m''| \neq |m|} P(|m| \to |m''|)\ , & |m| = |m'|\ ,
\end{cases}
\end{equation}
where $\Omega(|m|)$ is the degeneracy of state $|m|$ and $p_\text{MADE}(|m|)$ is the probability that the MADE generates a configuration of magnetization $|m|$, which can be computed in a time $\mathcal{O}(N)$ given the weights. The transition matrix for the signed magnetizations is similar and can be computed analogously.

\subsubsection{MADE and local Metropolis}

Finally, if we consider a global step followed by $k$ MCS, the transition matrix $\mathbb{P}_\text{M+L}$ will be simply given by the product
\begin{equation}
    \mathbb{P}_\text{M+L} = \mathbb{P}_\text{M}\mathbb{P}_\text{L}^{kN},
\end{equation}
where $\mathbb{P}_\text{L}$ and $\mathbb{P}_\text{M}$ are the transition matrices defined in the previous sections for the LMMC and MADE, respectively. In the following, we will always take $k = 1$ for simplicity.

In Sec. \ref{sec:results}, we compare the first passage times for the different methods to determine which procedure is the fastest to reach the target magnetization.

\section{Results}\label{sec:results}

In order to assess the relevance of adding LMMC to global NN-assisted moves, we consider the following setting, which corresponds to a single temperature jump in the GA procedure. 
We suppose that a NN has been trained (either perfectly or not) at the critical temperature $T_c$ of the model, where the spontaneous magnetization is still zero.
Then, we want to use the NN to perform
NN-assisted MC at a temperature $T<T_c$ below the critical temperature, at which
$m^* \neq 0$.
We compare the dynamics with global MADE moves only (indicated by M for MADE) with the dynamics with global MADE moves and LMMC (indicated by M+L for MADE+LMMC),
and compare the time it takes for the two dynamics to reach 
equilibrium, i.e. to reach $m^*$.

\subsection{Perfectly trained MADE}\label{sec:perfectly_trained}

We first consider a perfectly trained MADE, i.e., a MADE with weights given by the solution to Eq.~\eqref{eq:optimal_weights} at the critical temperature $\beta_c=1$ (we recall that these weights are non-zero at finite $N$).
We then evaluate the times required to reach the absolute equilibrium magnetization $|m^*|$ given by Eq.~\eqref{eq:m_CW} at inverse temperature $\beta\geq \beta_c$ (i.e. below the critical temperature) starting from zero magnetization,  and we consider the ratio
\begin{equation}\label{eq:R}
    R = \frac{2 \tau_\text{M+L}}{\tau_\text{M}} \ ,
\end{equation}
where $\tau_\text{M}$ and $\tau_\text{M+L}$ are the average first passage times for the two procedures and the factor two takes into account the fact that performing both a global move and a MCS takes approximately twice the number of operations as performing just a global move. Hence, $R>1$ indicates that the MADE by itself is more efficient than MADE+LMMC, while $R<1$ indicates that adding local moves is beneficial.

\begin{figure}[b]
    \centering
    \includegraphics[width=1\linewidth]{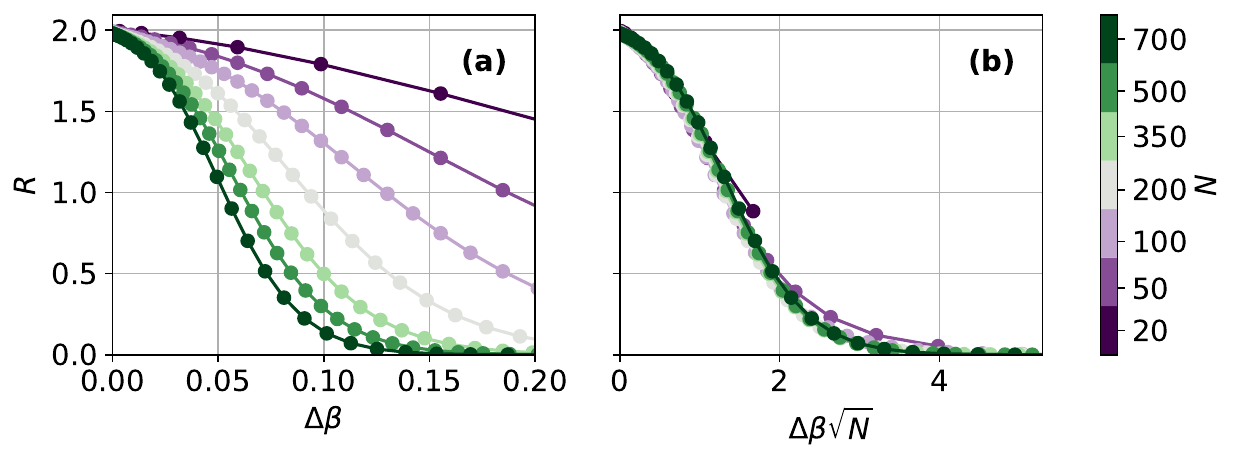}
    \caption{Ratio $R$ of first passage times as a function of $\Delta \beta = \beta -\beta_c$ (a) and of $\Delta \beta \sqrt{N}$ (b).}
    \label{fig:FPT_diffB}
\end{figure}

We first show in Fig. \ref{fig:FPT_diffB} the results for different $\beta$. The curves start at $R = 2$ for $\Delta \beta = \beta - \beta_c \simeq 0$, signaling that the perfectly trained NN is good enough to generate the target magnetization even on its own, without the need for local MC steps. Because the NN is good enough to generate configurations with the desired magnetization, adding LMMC on top of the global moves only adds to the computational time. Upon increasing $\Delta \beta$, as the machine is used at temperatures that are further away from the one at which it was trained, $R$ drops and it becomes increasingly necessary to add LMMC. The transition between these two regimes, however, moves at lower $\Delta \beta$ as $N$ increases. In particular, as shown in Fig. \ref{fig:FPT_diffB}b, the curves collapse when plotted as a function of $\Delta \beta \sqrt{N}$. Therefore, as long as the temperature step is chosen as $\Delta\beta = b/\sqrt{N}$ with $b$ small enough, a typical annealing schedule in practical applications, adding MC moves is actually not helpful. However, already for $b \gtrsim 1.5$ (which is not an uncommon choice), the LMMC clearly improves the performances of the MADE.
Moreover, this analysis does not take into account the (non-negligible) computational cost of training the MADE. Considering also the training time, the scenario changes, as we show in the following section.

\subsection{Partially trained MADE}\label{sec:non_perfectly_trained}

We now turn to the case in which the MADE is not already pre-trained and all the weights are initialized to zero.  This setting introduces a tradeoff in the training time: on the one hand, longer training is computationally more expensive; on the other hand, untrained weights make the performance worse (e.g. at initialization all weights are zero and the MADE simply extracts one configuration uniformly at random from the $2^N$ possible ones).
While it is unclear, \textit{a priori}, which is the optimal training time $t$,  we can track the evolution of the weights as a function of the number of training steps using Eq.~\eqref{eq:training_solution} and compute the average first passage times as in the previous section.
Following the discussion in Sec.~\ref{sec:training_dynamics}, we fix the learning rate $\eta = N^{-\frac{3}{2}}$. 

\begin{figure}
    \centering
    \includegraphics[width=1\linewidth]{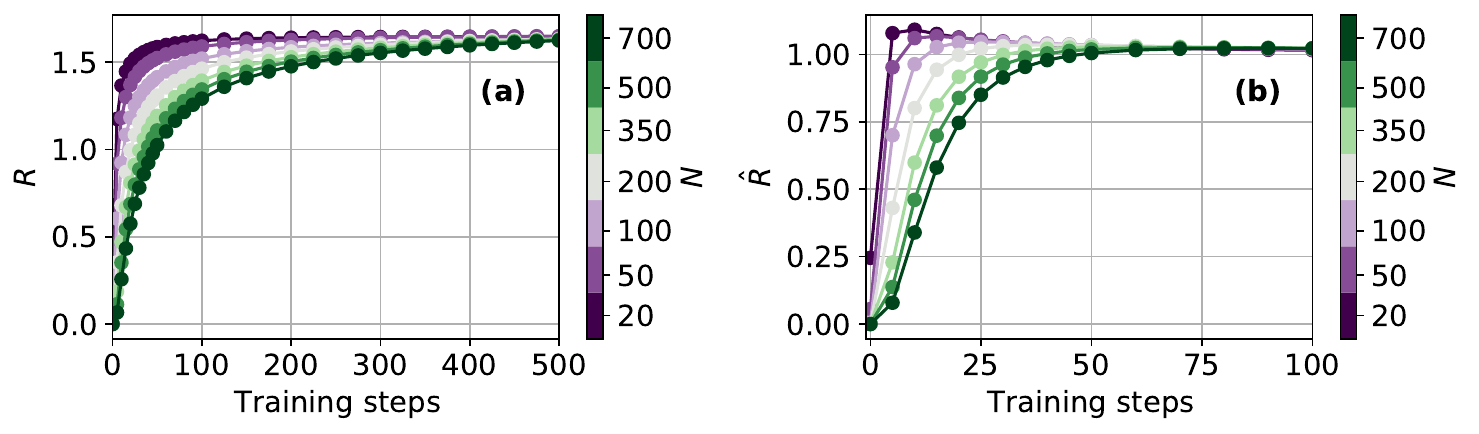}
    \caption{(a) Ratios of the average first passage time $R$ (not taking account training) and (b) $\hat{R}$ (taking account training) at $\Delta \beta = b/\sqrt{N}$ with $b = 0.5$. The learning rate is fixed to $\eta = N^{-\frac{3}{2}}$.}
    \label{fig:ratios_trainingtime}
\end{figure}

We plot $R$ at fixed $b=0.5$ as a function of the training time in Fig.~\ref{fig:ratios_trainingtime}a. 
From Fig.~\ref{fig:FPT_diffB}b, for $b=0.5$ we expect $R_\infty\sim 1.6$ at infinite training time, which is confirmed by Fig.~\ref{fig:ratios_trainingtime}a.
We see that, at finite training time, the performance of the MADE alone deteriorates with respect to MADE+LMMC, i.e. $R<R_\infty$. 
This phenomenon is even more evident Fig.~\ref{fig:ratios_trainingtime}b where we considered a modification of Eq.~\eqref{eq:R} that takes into account the training time $T_t$ (considering that each epoch takes $\sim N$ steps), i.e. 
the ratio
\begin{equation}
    \hat{R} = \frac{2 \tau_\text{M+L} + T_t}{\tau_\text{M} + T_t} \ .
\end{equation}
Performing LMMC helps, because it gives a huge hand when MADE is not trained enough, and even when MADE is good enough to be used alone, the addition of the training time adds an overhead that dominates over the additional cost of performing LMMC, so that the ratio remains always close to one or below.

We thus conclude that, if one has equilibrated at the critical temperature and wants
to equilibrate within one of the two states that appear just below it, MADE+LMMC is generically more efficient than MADE alone.

\subsection{Comparison with local Metropolis Monte Carlo}
\label{sec:comparison_vanilla}

\begin{figure}[t]
    \centering
    \includegraphics[width=1\linewidth]{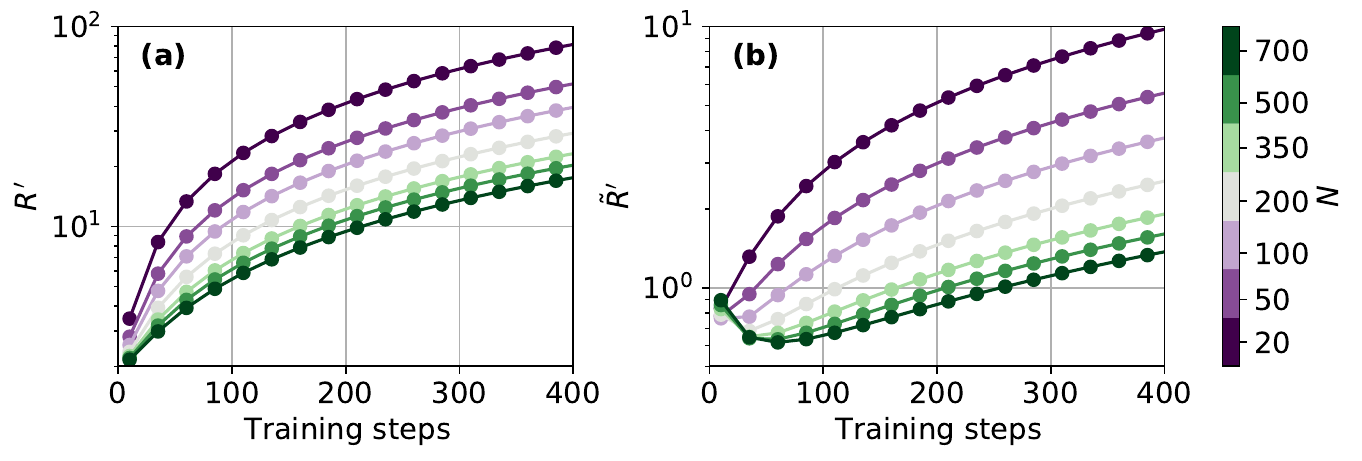}
    \caption{
    (a) Ratio $R'$ of the average first passage time between the MADE+LMMC (including training time) and the LMMC alone, for the absolute value of magnetization, as a function of training time for fixed $\Delta \beta = b/\sqrt{N}$ with $b = 2$.
    (b) Same plot for the ratio
    $\tilde{R}'$ for the signed value of the magnetization. 
    The learning rate is fixed to $\eta = N^{-\frac{3}{2}}$.
    }
    \label{fig:com_vanilla}
\end{figure}

Having assessed that MADE+LMMC is more efficient than MADE alone, one might wonder whether the MADE is needed at all. To answer this question,
we compare in the same setting the MADE+LMMC dynamics with that with LMMC only. 
The first passage time ratio
\begin{equation}
    R' = \frac{2\tau_\text{M+L} + T_t}{\tau_\text{L}}
\end{equation}
is shown in Fig.~\ref{fig:com_vanilla}a and
it turns out that simply using just LMMC performs better, i.e. we always observe that $R'>1$.

This is due to the fact that we are only considering the absolute value of the magnetization:
the addition of the MADE helps in exploring better the energy landscape by allowing sudden jumps between the two states that are present at $T<T_c$. 
This can be seen in Fig. \ref{fig:com_vanilla}b, in which we considered the ratio
\begin{equation}
    \tilde{R}' = \frac{2 \tilde{\tau}_\text{M+L} + T_t}{\tilde{\tau}_\text{L}}
\end{equation}
where $\tilde{\tau}_\text{L}$  and $\tilde{\tau}_\text{M+L}$ are the average first passage times of LMMC and MADE+LMMC for reaching the signed (positive) equilibrium magnetization starting from zero magnetization. In this case, we observe that $\tilde{R}' < 1$, indicating a better efficiency of MADE+LMMC, when $N$ is sufficiently large and the number of training steps is not too high (otherwise, the computational cost outweighs the gains from using MADE).
The LMMC alone has worse performance, because it can end up in the negative state and be stuck there for a long time. The addition of the MADE avoids this problem.
Note that the barrier for LMMC to jump from the negative to the positive state scales as $\exp(A N \Delta\beta^2) = \exp(A b^2)$, hence it remains finite with the chosen scaling of $\Delta\beta = b/\sqrt{N}$. When instead $\beta-\beta_c=O(1)$, the LMMC needs a time scaling exponentially in $N$, and the gain from using the MADE becomes even more visible.
\new{We expect a similar behavior to hold in the whole ferromagnetic region.}

\section{Conclusions}\label{sec:conclusions}

In this work, we were able to fully describe a generative autoregressive architecture, the shallow MADE, for the study of the Curie-Weiss model at finite size $N$. We first characterized the problem in terms of the optimal couplings that can be found in order to approximate the Curie-Weiss distribution. We were then able to describe how these couplings are learned during the training process. Interestingly, we found that the system undergoes a critical slowing down in the learning characterized by the same behavior as typical local dynamics. Further work is needed to test whether this is a general result or a peculiarity of the model.

We were then able to use these results to benchmark the model performances with and without additional local Monte Carlo steps in the Global Annealing procedure. We found that using the perfectly trained architecture renders additional local Monte Carlo steps unnecessary, as long as a suitable annealing schedule is chosen (i.e. with small enough $b =\Delta\beta \sqrt{N}$). However, since the cost of training the architecture increases with the model size, one has to resort to using an imperfectly trained machine, which in turn benefits from using additional local Monte Carlo steps. However, we verified that the NN-assisted procedure is actually able to outperform that with local moves only because it allows jumping between distinct states that form below the critical temperature, thus showing the improvement coming from the usage of the architecture. In summary, the NN role is mostly to allow for efficient sampling of distinct states, while the local moves are mostly needed to efficiently sample within a state.

Based on these results we predict that, in practical applications, the use of generative autoregressive NN is helpful to better simulate the systems of interest whenever more than one state is present. 
However, since in practice one has to compromise between the accuracy and time of training, the addition of Monte Carlo steps will actually improve the results. 
Further numerical tests of these ideas are left for future work.

\section*{Code availability}
The code used in this paper is available at the GitHub repository \cite{MonteCarloGA_CW}.

\section*{Acknowledgments}
We thank Giulio Biroli, Marylou Gabri\'e and Guilhem Semerjian for useful discussions.
The research has received financial support from the “National Centre for HPC, Big Data and Quantum Computing - HPC”, Project CN\_00000013, CUP B83C22002940006, NRP Mission 4 Component 2 Investment 1.4,  Funded by the European Union - NextGenerationEU. Author LMDB acknowledges funding from the Bando Ricerca Scientifica 2024 - \textit{Avvio alla Ricerca} (D.R. No. 1179/2024) of Sapienza Università di Roma, project B83C24005280001 – MaLeDiSSi. We acknowledge support from the computational infrastructure DARIAH.IT, PON Project code PIR01\_00022, National Research Council of Italy.

\bibliographystyle{unsrt}
\bibliography{references}

\pagebreak

\onecolumngrid

\appendix

\setcounter{figure}{0}

\pagebreak

\pagenumbering{gobble}

\end{document}